# The Kauffman Constraint Coefficients $K_\omega(n_0, n_1, \ldots, n_{\beta-1})$

Kenneth A. Griggs
Phoenix project
Washington, DC 20009
Kenneth.A.Griggs@gmail.com

## I. Introduction

In Reference [1], Louis H. Kauffman presents an *Algebra of Constraints*.[1] Its purpose is to explicitly define the algebraic conditions that best align non-commutative operators with their corresponding classical variables in the usual context of continuum calculus.

In creating this *classical variable* (CV) to *non-commutating operator* (NCO) correspondence, two assumptions are made:

(1) there exists an exact mapping between CVs and NCOs of the form

**(I-1)**

$$\begin{gathered} \textit{classical variable} \leftrightarrow \textit{Non-Commutating Operator} \\ x_1^n \leftrightarrow X_1^n \\ x_2^n \leftrightarrow X_2^n \\ \vdots \\ x_\beta^n \leftrightarrow X_\beta^n \end{gathered}$$

where $n$ is a positive integer, and

(2) there exists an exact mapping between linear combinations of CVs and linear combinations of NCOs of the form

**(I-2)**

$$\begin{gathered} \textit{classical variables} \leftrightarrow \textit{Non-Commutating Operators} \\ \left(\sum_{\alpha=1}^{\beta} x_\alpha\right)^n \leftrightarrow \left(\sum_{\alpha=1}^{\beta} X_\alpha\right)^n \end{gathered}$$

These two requirements establish constraint conditions on the NCOs that take the form of *Symmetrizers*; namely,

**(I-3)**

$$\left\{X_1 X_2 \cdots X_\beta\right\} \leftrightarrow \left(\frac{1}{n!}\right) \sum_{\sigma \in S_n} \left(X_{\sigma_1} X_{\sigma_2} \cdots X_{\sigma_n}\right)$$

where we are summing over all $n$-permutations of the permutation group $S_n$.[a]

[a] Kauffman uses {} brackets to designate *Symmetrizers* as opposed to the [] which designate *Commutators*. We will only use his Symmetrization convention in equation (I-3). In all cases that follow, curly brackets {} designate simple parenthesis.

These mappings enable Kauffman to align classical variable differential formulas with their corresponding non-commutative operator differential forms. In fact, by defining the first temporal derivative of a variable $\theta$ with respect to time

**(I-4)**

$$\theta^{(1)} = h\theta = h^{(0)}\theta$$

Kauffman provides a general algorithm for obtaining a $\beta$-order time-derivative $\theta^{(\beta)}$ in a recursive fashion. Namely, the next derivative is obtained from the previous by applying the product rule for differentiation and using the (I-4) identity. By programming this algorithm into Mathematica™, Kauffman is able to calculate the first nine levels of derivatives. However, Kauffman notes that

> "*The structure of the coefficients in this recursion is unknown territory...To penetrate the full algebra of constraints we need to understand the structure of these derivatives and their corresponding non-commutative symmetrizations.*"[b]

While Kauffman employs computational differentiation using Mathematica™, we present below a closed-form expression for the *Kauffman Constraint Coefficient* $K_\omega(n_0, n_1, n_2, \cdots, n_{\beta-1})$ both as a solution to his challenge and to foster a greater understanding of the full *Algebra of Constraints*.

## II. Solution

The *Algebra of Constraints* can be described combinatorically in terms of its structures. We begin by defining a $\beta$-order time-derivative $\theta^{(\beta)}$ in terms of three structures: the first is the original variable $\theta$; the second is a set of $\beta$-derivatives $\{H^{(0)}, H^{(1)}, \ldots, H^{(\alpha)}, \ldots, H^{(\beta-2)}, H^{(\beta-1)}\}$; and the third is a set defining the multiplicity of those $\beta$-derivatives $\{n_0, n_1, \ldots, n_\alpha, \ldots, n_{\beta-2}, n_{\beta-1}\}$. From these structures, two generalized functions are defined: firstly, there is the *ω-Elemental* $E_\omega$, which is a product of the $\beta$-derivatives with their $n_\alpha$ multiplicities; secondly, there is the *Kauffman Constraint Coefficient* $K_\omega$, which is a count of the ω-Elementals.

With these two functions, Kauffman's derivative series $\theta^{(\beta)}$ takes the general form

**(II-1)**

$$\theta^{(\beta)} = \sum_{\omega=1}^{N_\beta} K_\omega(n_0, n_1, \cdots, n_{\beta-1}) E_\omega(n_0, H^{(0)}, n_1, H^{(1)}, \cdots, n_{\beta-1}, H^{(\beta-1)})$$

where every Elemental $E_\omega$ of the derivative series is explicitly given by[c]

---

[b] Ibid., p. 31.

[c] Because the Coefficient $\boldsymbol{K_\omega}$ is independent of the existence of the time functional $\theta = \theta^{(0)} = \boldsymbol{T}$ (where T is used in Kauffman's paper) for every Elemental of the derivative $\theta^{(\beta)}$ expansion, it is not displayed in the $\boldsymbol{E_\omega}$ Elementals.

**(II-2)**

$$E_\omega\left(n_0, H^{(0)}, n_1, H^{(1)}, \cdots, n_{\beta-1}, H^{(\beta-1)}\right) = \left\{ \prod_{\alpha=0}^{\beta-1} \left(H^{(\alpha)}\right)^{n_\alpha} \right\}_\omega$$

$$= \left\{ \left(H^{(0)}\right)^{n_0} \left(H^{(1)}\right)^{n_1} \cdots \left(H^{(\alpha)}\right)^{n_\alpha} \cdots \left(H^{(\beta-2)}\right)^{n_{\beta-2}} \left(H^{(\beta-1)}\right)^{n_{\beta-1}} \right\}_\omega$$

with

$\beta \equiv$ the number of derivatives of $\theta$ with respect to the Hamiltonian $T$ , $\theta^{(\beta)}$;
$N_\beta \equiv$ the total number of distinct Elementals $E_\omega$, with $1 \le \omega \le N_\beta$ , of the derivative series;
$n_\alpha \equiv$ the exponent and number of $\alpha$ -derivatives, $H^{(\alpha) n_\alpha}$ ;
$K_\omega\left(n_0, n_1, \cdots, n_{\beta-1}\right) \equiv$ the *ω-Kauffman Constraint Coefficient* for the ω-Elemental of $\theta^{(\beta)}$

The key to this construction is that all $\alpha$ -derivatives $H^{(\alpha)}$, where $0 \le \alpha \le \beta - 1$, are present for every Elemental $E_\omega$ of the series $\theta^{(\beta)}$. As such, the form of the Elemental is encoded in the set of exponents $n_\alpha$. Determining the correct ω-sets $\left\{n_0, n_1, \cdots, n_{\beta-1}\right\}_\omega$ of exponents generates the correct $E_\omega$ elementals and their corresponding *Kauffman Constraint Coefficients* $K_\omega$.[d]

Because of this, the *Kauffman Constraint Coefficient* for each $E_\omega$ Elemental is purely a function of the $n_\alpha$ exponents instead of the $\alpha$ -derivatives $H^{(\alpha)}$. As such, the *Kauffman Constraint Coefficient* is defined as

**(II-3)**

$$K_\omega\left(n_0, n_1, \cdots, n_{\beta-1}\right) \equiv \frac{\beta!}{\prod_{\alpha=0}^{\beta-1} \left\{ n_\alpha! [(\alpha+1)!]^{n_\alpha} \right\}}$$

## III. Redefining the Kauffman Recursion Relation for $\theta^{(\beta)}$

As explained in the *Introduction*, Kauffman generates each time derivative $\theta^{(\beta)}$ via a commutative recursion relation with all smaller derivatives

**(III-1)**

$$\left\{ \theta^{(0)}, \theta^{(1)}, \ldots, \theta^{(\beta-1)} \right\}$$

We have derived a new recursion relation that does not depend on using the product rule for differentiation but instead involves only summations and substitutions.[e] Namely,

**(III-2)**

$$\theta^{(\beta)} = (\beta-1)! \sum_{\alpha=1}^{\beta} \left\{ \frac{H^{(\alpha-1)} \theta^{(\beta-\alpha)}}{(\alpha-1)!\,(\beta-\alpha)!} \right\}$$

---

[d] Calculating the $\boldsymbol{E_\omega}$ Elementals can be performed as an iterative optimization problem through recursive differentiation and application of the product rule of differentiation or alternatively by *Combinatoric* methods which will be explored in an upcoming paper.

[e] This technique may provide a cleaner, more efficient algorithm for Mathematica™.

As a demonstration, we explicitly generate the first 4 such derivatives in the following Table III-1.

**Table III-1: New Recursive Calculation of $\theta^{(\beta)}$**

| $\theta^{(\beta)}$ | $=(\beta-1)!\sum_{\alpha=1}^{\beta}\left\{\frac{H^{(\alpha-1)}\theta^{(\beta-\alpha)}}{(\alpha-1)!(\beta-\alpha)!}\right\}$ | **Recursive Calculation** | **Result** |
|---|---|---|---|
| $\theta^{(1)}$ | $=(0)!\sum_{\alpha=1}^{1}\left\{\frac{H^{(\alpha-1)}\theta^{(1-\alpha)}}{(\alpha-1)!(1-\alpha)!}\right\}$ | $=\frac{H^{(0)}\theta^{(0)}}{(0)!(0)!}$ | $H^{(0)}\theta^{(0)}=H\theta$ |
| $\theta^{(2)}$ | $=(1)!\sum_{\alpha=1}^{2}\left\{\frac{H^{(\alpha-1)}\theta^{(2-\alpha)}}{(\alpha-1)!(2-\alpha)!}\right\}$ | $=\frac{H^{(0)}\theta^{(1)}}{(0)!(1)!}+\frac{H^{(1)}\theta^{(0)}}{(1)!(0)!}=H^{(0)}\theta^{(1)}+H^{(1)}\theta^{(0)}$ <br> $=H(H\theta)+H^{(1)}\theta$ | $H^2\theta+H^{(1)}\theta$ |
| $\theta^{(3)}$ | $=(2)!\sum_{\alpha=1}^{3}\left\{\frac{H^{(\alpha-1)}\theta^{(3-\alpha)}}{(\alpha-1)!(3-\alpha)!}\right\}$ | $=(2)!\left\{\frac{H^{(0)}\theta^{(2)}}{(0)!(2)!}+\frac{H^{(1)}\theta^{(1)}}{(1)!(1)!}+\frac{H^{(2)}\theta^{(0)}}{(2)!(0)!}\right\}$ <br> $=(2)!\left\{\frac{H^{(0)}\theta^{(2)}}{2}+\frac{H^{(1)}\theta^{(1)}}{1}+\frac{H^{(2)}\theta^{(0)}}{2}\right\}$ <br> $=H\left(H^2\theta+H^{(1)}\theta\right)+2H^{(1)}(H\theta)+H^{(2)}\theta$ | $H^3\theta+3HH^{(1)}\theta+H^{(2)}\theta$ |
| $\theta^{(4)}$ | $=(3)!\sum_{\alpha=1}^{4}\left\{\frac{H^{(\alpha-1)}\theta^{(4-\alpha)}}{(\alpha-1)!(4-\alpha)!}\right\}$ | $=(3)!\left\{\frac{H^{(0)}\theta^{(3)}}{(0)!(3)!}+\frac{H^{(1)}\theta^{(2)}}{(1)!(2)!}+\frac{H^{(2)}\theta^{(1)}}{(2)!(1)!}+\frac{H^{(3)}\theta^{(0)}}{(3)!(0)!}\right\}$ <br> $=(3)!\left\{\frac{H^{(0)}\theta^{(3)}}{6}+\frac{H^{(1)}\theta^{(2)}}{2}+\frac{H^{(2)}\theta^{(1)}}{2}+\frac{H^{(3)}\theta^{(0)}}{6}\right\}$ <br> $=H\left(H^3\theta+3HH^{(1)}\theta+H^{(2)}\theta\right)$ <br> $+3H^{(1)}\left(H^2\theta+H^{(1)}\theta\right)$ <br> $+3H^{(2)}(H\theta)+H^{(3)}\theta$ | $H^4\theta+6H^2H^{(1)}\theta$ <br> $+4HH^{(2)}\theta+3H^{(1)2}\theta$ <br> $+H^{(3)}\theta$ |

## IV. Generating $K_\omega$ for $\theta^{(9)}$

As discussed in the *Introduction*, Kauffman calculates the first nine levels of derivatives via Mathematica™.[f] We now present the calculation for those same coefficients using Equation (II-3). In so doing, we demonstrate how properly to use the Elementals $(E_1, E_2, E_3, \cdots, E_{30})$ corresponding to the $\theta^{(9)}$ derivative with the *Kauffman Constraint Coefficient* function to obtain the 30 coefficients $(K_1, K_2, K_3, \cdots, K_{30})$.

**Table IV-1: Kauffman Constraint Coefficient for $\theta^{(9)}$**

| $\omega$ | Mathematica™ | $E_\omega$ | $K_\omega(n_0,n_1,n_2,n_3,\ldots,n_8)$ | Calculation |
|---|---|---|---|---|
| 1 | $1\cdot H^{(0)9}T$ | $H^{(0)9}$ | $K_1(9,0,0,0,0,0,0,0,0)$ | $\frac{\beta!}{n_0!(1!)^{n_0}}=\frac{9!}{9!(1!)^9}=1$ |
| 2 | $36\cdot H^{(0)7}TH^{(1)1}$ | $H^{(0)7}H^{(1)1}$ | $K_2(7,1,0,0,0,0,0,0,0)$ | $\frac{\beta!}{\left(n_0!(1!)^{n_0}\right)\left(n_1!(2!)^{n_1}\right)}$ <br> $=\frac{9!}{7!\,2}=\frac{9!}{10080}=36$ |
| 3 | $378\cdot H^{(0)5}TH^{(1)2}$ | $H^{(0)5}H^{(1)2}$ | $K_3(5,2,0,0,0,0,0,0,0)$ | $\frac{\beta!}{\left(n_0!(1!)^{n_0}\right)\left(n_1!(2!)^{n_1}\right)}$ <br> $=\frac{9!}{5!\,8}=\frac{9!}{960}=378$ |

[f] Ibid., p. 32.

| $\omega$ | Mathematica™ | $E_\omega$ | $K_\omega(n_0,n_1,n_2,n_3,\ldots,n_8)$ | Calculation |
|---|---|---|---|---|
| 4 | $1260 \cdot H^{(0)3}TH^{(1)3}$ | $H^{(0)3}H^{(1)3}$ | $K_4(3,3,0,0,0,0,0,0,0)$ | $\frac{\beta!}{(n_0!(1!)^{n_0})(n_1!(2!)^{n_1})} = \frac{9!}{3!\,3!\,8} = \frac{9!}{288} = 1260$ |
| 5 | $945 \cdot H^{(0)1}TH^{(1)4}$ | $H^{(0)1}H^{(1)4}$ | $K_5(1,4,0,0,0,0,0,0,0)$ | $\frac{\beta!}{(n_0!(1!)^{n_0})(n_1!(2!)^{n_1})} = \frac{9!}{1!\,4!\,16} = \frac{9!}{384} = 945$ |
| 6 | $84 \cdot H^{(0)6}TH^{(2)1}$ | $H^{(0)6}H^{(2)1}$ | $K_6(6,0,1,0,0,0,0,0,0)$ | $\frac{\beta!}{(n_0!(1!)^{n_0})(n_2!(3!)^{n_2})} = \frac{9!}{6!\,3!} = \frac{9!}{4320} = 84$ |
| 7 | $1260 \cdot H^{(0)4}TH^{(1)1}H^{(2)1}$ | $H^{(0)4}H^{(1)1}H^{(2)1}$ | $K_7(4,1,1,0,0,0,0,0,0)$ | $\frac{\beta!}{(n_0!(1!)^{n_0})(n_1!(2!)^{n_1})(n_2!(3!)^{n_2})} = \frac{9!}{4!\,3!\,2!} = \frac{9!}{288} = 1260$ |
| 8 | $3780 \cdot H^{(0)2}TH^{(1)2}H^{(2)1}$ | $H^{(0)2}H^{(1)2}H^{(2)1}$ | $K_8(2,2,1,0,0,0,0,0,0)$ | $\frac{\beta!}{(n_0!(1!)^{n_0})(n_1!(2!)^{n_1})(n_2!(3!)^{n_2})} = \frac{9!}{2!\,2!\,(2!)^2\,3!} = \frac{9!}{96} = 3780$ |
| 9 | $1260 \cdot TH^{(1)3}H^{(2)1}$ | $H^{(1)3}H^{(2)1}$ | $K_9(0,3,1,0,0,0,0,0,0)$ | $\frac{\beta!}{(n_1!(2!)^{n_1})(n_2!(3!)^{n_2})} = \frac{9!}{3!\,(2!)^3\,3!} = \frac{9!}{288} = 1260$ |
| 10 | $840 \cdot H^{(0)3}TH^{(2)2}$ | $H^{(0)3}H^{(2)2}$ | $K_{10}(3,0,2,0,0,0,0,0,0)$ | $\frac{\beta!}{(n_0!(1!)^{n_0})(n_2!(3!)^{n_2})} = \frac{9!}{3!\,2!\,(3!)^2} = \frac{9!}{432} = 840$ |
| 11 | $2520 \cdot H^{(0)1}TH^{(1)1}H^{(2)2}$ | $H^{(0)1}H^{(1)1}H^{(2)2}$ | $K_{11}(1,1,2,0,0,0,0,0,0)$ | $\frac{\beta!}{(n_0!(1!)^{n_0})(n_1!(2!)^{n_1})(n_2!(3!)^{n_2})} = \frac{9!}{1!\,2!\,2!\,(3!)^2} = \frac{9!}{144} = 2520$ |
| 12 | $280 \cdot TH^{(2)3}$ | $H^{(2)3}$ | $K_{12}(0,0,3,0,0,0,0,0,0)$ | $\frac{\beta!}{(n_2!(3!)^{n_2})} = \frac{9!}{3!\,(3!)^3} = \frac{9!}{1296} = 280$ |
| 13 | $126 \cdot H^{(0)5}TH^{(3)1}$ | $H^{(0)5}H^{(3)1}$ | $K_{13}(5,0,0,1,0,0,0,0,0)$ | $\frac{\beta!}{(n_0!(1!)^{n_0})(n_3!(4!)^{n_3})} = \frac{9!}{5!\,4!} = \frac{9!}{2880} = 126$ |
| 14 | $1260 \cdot H^{(0)3}TH^{(1)1}H^{(3)1}$ | $H^{(0)3}H^{(1)1}H^{(3)1}$ | $K_{14}(3,1,0,1,0,0,0,0,0)$ | $\frac{\beta!}{(n_0!(1!)^{n_0})(n_1!(2!)^{n_1})(n_3!(4!)^{n_3})} = \frac{9!}{3!\,2!\,4!} = \frac{9!}{288} = 1260$ |
| 15 | $1890 \cdot H^{(0)1}TH^{(1)2}H^{(3)1}$ | $H^{(0)1}H^{(1)2}H^{(3)1}$ | $K_{15}(1,2,0,1,0,0,0,0,0)$ | $\frac{\beta!}{(n_0!(1!)^{n_0})(n_1!(2!)^{n_1})(n_3!(4!)^{n_3})} = \frac{9!}{1!\,(2!)^3\,4!} = \frac{9!}{192} = 1890$ |
| 16 | $1260 \cdot H^{(0)2}TH^{(2)1}H^{(3)1}$ | $H^{(0)2}H^{(2)1}H^{(3)1}$ | $K_{16}(2,0,1,1,0,0,0,0,0)$ | $\frac{\beta!}{(n_0!(1!)^{n_0})(n_2!(3!)^{n_2})(n_3!(4!)^{n_3})} = \frac{9!}{2!\,3!\,4!} = \frac{9!}{288} = 1260$ |
| 17 | $1260 \cdot TH^{(1)1}H^{(2)1}H^{(3)1}$ | $H^{(1)1}H^{(2)1}H^{(3)1}$ | $K_{17}(0,1,1,1,0,0,0,0,0)$ | $\frac{\beta!}{(n_1!(2!)^{n_1})(n_2!(3!)^{n_2})(n_3!(4!)^{n_3})} = \frac{9!}{2!\,3!\,4!} = \frac{9!}{288} = 1260$ |

| $\omega$ | Mathematica™ | $E_\omega$ | $K_\omega(n_0,n_1,n_2,n_3,\ldots,n_8)$ | Calculation |
|---|---|---|---|---|
| 18 | $315\cdot H^{(0)1}TH^{(3)2}$ | $H^{(0)1}H^{(3)2}$ | $K_{18}(1,0,0,2,0,0,0,0,0)$ | $\frac{\beta!}{(n_0!(1!)^{n_0})(n_3!(4!)^{n_3})} = \frac{9!}{1!\,2!(4!)^2} = \frac{9!}{1152} = 315$ |
| 19 | $126\cdot H^{(0)4}TH^{(4)1}$ | $H^{(0)4}H^{(4)1}$ | $K_{19}(4,0,0,0,1,0,0,0,0)$ | $\frac{\beta!}{(n_0!(1!)^{n_0})(n_4!(5!)^{n_4})} = \frac{9!}{4!\,5!} = \frac{9!}{2880} = 126$ |
| 20 | $756\cdot H^{(0)2}TH^{(1)1}H^{(4)1}$ | $H^{(0)2}H^{(1)1}H^{(4)1}$ | $K_{20}(2,1,0,0,1,0,0,0,0)$ | $\frac{\beta!}{(n_0!(1!)^{n_0})(n_1!(2!)^{n_1})(n_4!(5!)^{n_4})} = \frac{9!}{2!\,2!\,5!} = \frac{9!}{480} = 756$ |
| 21 | $378\cdot TH^{(1)2}H^{(4)1}$ | $H^{(1)2}H^{(4)1}$ | $K_{21}(0,2,0,0,1,0,0,0,0)$ | $\frac{\beta!}{(n_1!(2!)^{n_1})(n_4!(5!)^{n_4})} = \frac{9!}{(2!)^3 5!} = \frac{9!}{960} = 378$ |
| 22 | $504\cdot H^{(0)1}TH^{(2)1}H^{(4)1}$ | $H^{(0)1}H^{(2)1}H^{(4)1}$ | $K_{22}(1,0,1,0,1,0,0,0,0)$ | $\frac{\beta!}{(n_0!(1!)^{n_0})(n_2!(3!)^{n_2})(n_4!(5!)^{n_4})} = \frac{9!}{1!\,3!\,5!} = \frac{9!}{720} = 504$ |
| 23 | $126\cdot TH^{(3)1}H^{(4)1}$ | $H^{(3)1}H^{(4)1}$ | $K_{23}(0,0,0,1,1,0,0,0,0)$ | $\frac{\beta!}{(n_3!(4!)^{n_3})(n_4!(5!)^{n_4})} = \frac{9!}{4!\,5!} = \frac{9!}{2880} = 126$ |
| 24 | $84\cdot H^{(0)3}TH^{(5)1}$ | $H^{(0)3}H^{(5)1}$ | $K_{24}(3,0,0,0,0,1,0,0,0)$ | $\frac{\beta!}{(n_0!(1!)^{n_0})(n_5!(6!)^{n_5})} = \frac{9!}{3!\,6!} = \frac{9!}{4320} = 84$ |
| 25 | $252\cdot H^{(0)1}TH^{(1)1}H^{(5)1}$ | $H^{(0)1}H^{(1)1}H^{(5)1}$ | $K_{25}(1,1,0,0,0,1,0,0,0)$ | $\frac{\beta!}{(n_0!(1!)^{n_0})(n_1!(2!)^{n_1})(n_5!(6!)^{n_5})} = \frac{9!}{1!\,2!\,6!} = \frac{9!}{1440} = 252$ |
| 26 | $84\cdot TH^{(2)1}H^{(5)1}$ | $H^{(2)1}H^{(5)1}$ | $K_{26}(0,0,1,0,0,1,0,0,0)$ | $\frac{\beta!}{(n_2!(3!)^{n_2})(n_5!(6!)^{n_5})} = \frac{9!}{3!\,6!} = \frac{9!}{4320} = 84$ |
| 27 | $36\cdot H^{(0)2}TH^{(6)1}$ | $H^{(0)2}H^{(6)1}$ | $K_{27}(2,0,0,0,0,0,1,0,0)$ | $\frac{\beta!}{(n_0!(1!)^{n_0})(n_6!(7!)^{n_6})} = \frac{9!}{2!\,7!} = \frac{9!}{10080} = 36$ |
| 28 | $36\cdot TH^{(1)1}H^{(6)1}$ | $H^{(1)1}H^{(6)1}$ | $K_{28}(0,1,0,0,0,0,1,0,0)$ | $\frac{\beta!}{(n_1!(2!)^{n_1})(n_6!(7!)^{n_6})} = \frac{9!}{2!\,7!} = \frac{9!}{10080} = 36$ |
| 29 | $9\cdot H^{(0)1}TH^{(7)1}$ | $H^{(0)1}H^{(7)1}$ | $K_{29}(1,0,0,0,0,0,0,1,0)$ | $\frac{\beta!}{(n_0!(1!)^{n_0})(n_7!(8!)^{n_7})} = \frac{9!}{1!\,8!} = \frac{9!}{40320} = 9$ |
| 30 | $1\cdot TH^{(8)1}$ | $H^{(8)1}$ | $K_{30}(0,0,0,0,0,0,0,0,1)$ | $\frac{\beta!}{(n_8!(9!)^{n_8})} = \frac{9!}{9!} = 1$ |

## V. Generating $K_7$ and $K_{33}$ for $\theta^{(12)}$

We extend our application of the *Kauffman Constraint Coefficient* $K_\omega$ to the slightly more difficult $\theta^{(12)}$ derivative. In so doing, we will only calculate the coefficients for two Elementals, namely,

**(V-1)**

$$E_7 = H^{(0)3}H^{(2)3}$$
$$E_{33} = H^{(0)1}H^{(4)1}H^{(5)1}$$

The corresponding *Kauffman Constraint Coefficients* are

**(V-2)**

$$K_7(3,0,3,0,0,0,0,0,0,0,0,0,0) = \frac{\beta!}{(n_0!(1!)^{n_0})(n_2!(3!)^{n_2})} = \frac{12!}{(3!)(3!(3!)^3)} = \frac{479{,}001{,}600}{7776} = 61{,}600$$
$$K_{33}(1,0,0,0,1,1,0,0,0,0,0,0,0) = \frac{\beta!}{(n_0!(1!)^{n_0})(n_4!(5!)^{n_4})(n_5!(6!)^{n_5})} = \frac{12!}{(1!(1!)^1)(1!(5!)^1)(1!(6!)^1)} = \frac{479{,}001{,}600}{86{,}400} = 5{,}544$$

The reader should find these calculations using Mathematica™, or many other software programs, computationally reasonable.

## VI. The $\theta^{(40)}$ Challenge

To further demonstrate the usefulness of the *Kauffman Constraint Coefficient*, a quick calculation can be made that would otherwise push the computational limits of Mathematica™ on many non-supercomputers. Namely, the reader may find it challenging to find the coefficient of the Elemental

**(VI-1)**

$$E_\omega = H^{(0)1}H^{(1)1}H^{(36)1}$$

of the fortieth time derivative $\theta^{(40)}$. The predicted *Kauffman Constraint Coefficient* is

**(VI-2)**

$$K_\omega(1,1,0,0,0,0,0,0,0,0,0,0,0,0,0,0,0,0,0,0,0,0,0,0,0,0,0,0,0,0,0,0,0,0,0,0,1,0,0,0)$$
$$= \frac{\beta!}{(n_0!(1!)^{n_0})(n_1!(2!)^{n_1})(n_{36}!(37!)^{n_{36}})} = \frac{40!}{1!\,2!\,37!} = \frac{38*39*40}{2} = 29{,}640$$

## VII. Conclusion

Because this paper is not designed to recount the full scope of *Non-Commutative Worlds*[g] as explicitly described by Louis H. Kauffman, we invite the reader to become more fully engaged in his revolutionary work. As such, we have presented only a summary of Kauffman's mathematics and expressed motivations for creating an *Algebra of Constraints*. From this, we answered an outstanding question of the algebra: *What is the structure of the coefficients in the Kauffman Recursion Relation for* $\theta^{(\beta)}$*?*

---

[g] Ibid., pps. 1-34.

The *Kauffman Constraint Coefficients* $K_\omega$ and their corresponding *Elementals* $E_\omega$ are presented as solutions to the construction of the $\theta^{(\beta)}$ derivative. Additionally, a new recursion relation is provided that requires only operational substitutions and summations; this algorithmically simplifies Kauffman's original technique. To demonstrate $K_\omega$, we generate the 30 $K_\omega$ *Coefficients* from the corresponding *Elementals* $E_\omega$ for $\theta^{(9)}$ and find that our results are in complete agreement with Kauffman's Mathematica™ solutions. We further present a calculation of two coefficients for the $\theta^{(12)}$ derivative and invite readers to use Mathematica™ or any other means to calculate and verify our results. Finally, we present a challenging calculation for a coefficient of the $\theta^{(40)}$ derivative series; owing to the vast numbers of permutations involved, a Mathematica™ approach may require substantial computer resources to obtain the solution in a reasonable time.

These formula, calculations and techniques for the *Kauffman Constraint Coefficients* $K_\omega$ and their corresponding *Elementals* $E_\omega$ are presented to enable the further development and discussion of Kauffman's emerging *Algebra of Constraints*.

---

## VII. References

[1] Louis H. Kauffman [2011], Non-Commutative Worlds – Classical Constraints, Relativity and the Bianchi Identity, ArXiv:1109.1085v1 [math-ph], September 06, 2011, 23-26, 30-32.